\documentclass[aps,prb,preprint,superscriptaddress]{revtex4-1}
\usepackage{graphicx,amsmath,amsfonts,mathrsfs}
\newcommand{\C}{$^{\circ}$C}
\begin{document}

\title{Experimental and first-principles studies of magnetism and magnetoelectric effect in Co$_4$Nb$_2$O$_9$ and Co$_4$Ta$_2$O$_9$}
\author{I. V. Solovyev}
\email{SOLOVYEV.Igor@nims.go.jp}
\affiliation{National Institute for Materials
Science, 1-1 Namiki, Tsukuba, Ibaraki 305-0044, Japan}
\affiliation{Department of Theoretical Physics and Applied Mathematics, Ural Federal
University, Mira str. 19, 620002 Ekaterinburg, Russia}
\author{T. V. Kolodiazhnyi}
\affiliation{National Institute for Materials
Science, 1-1 Namiki, Tsukuba, Ibaraki 305-0044, Japan}
\date{\today}

\begin{abstract}
We report results of joint experimental and theoretical studies on magnetoelectric (ME) compounds Co$_4$Nb$_2$O$_9$ and Co$_4$Ta$_2$O$_9$.
On the experimental side, we present results of the magnetization and dielectric permittivity measurements in the magnetic field.
On the theoretical side, we construct the low-energy Hubbard-type model for the magnetically active Co$3d$ bands in the Wannier basis,
using the input of first-principles electronic structure calculations, solve this model in the mean-field Hartree-Fock approximation,
and evaluate the electric polarization in terms of the Berry phase theory. Both experimental
and theoretical results suggest that Co$_4$Ta$_2$O$_9$ is magnetically softer than Co$_4$Nb$_2$O$_9$. Therefore, it is reasonable to expect that the antiferromagnetic structure of Co$_4$Ta$_2$O$_9$ can be easier deformed by the external magnetic field, yielding larger polarization. This trend is indeed reproduced by our
theoretical calculations, but does not seem to be consistent with the experimental behavior of the polarization and dielectric permittivity.
Thus, we suggest that there should be a hidden mechanism controlling the ME coupling in these compounds,
probably related to the
magnetic striction or a spontaneous change of the magnetic structure, which breaks the inversion symmetry. Furthermore,
we argue that unlike in other ME systems (e.g. Cr$_2$O$_3$), in Co$_4$Nb$_2$O$_9$ and Co$_4$Ta$_2$O$_9$ there are
two crystallographic sublattices, which contribute to the ME effect. These contributions are
found to be
of the opposite sign and tend to compensate
each other. The latter mechanism can be also used to control and reverse the electric polarization in these compounds.
\end{abstract}

\maketitle




\section{\label{sec:Intro} Introduction}
The exploration of magnetoelectric (ME) effect -- the phenomenon, when the magnetization (electric polarization) can be induced by the external
electric (magnetic) field -- has attracted a great deal of attention, due to its potential applicability in the new generation of
multifunctional electronic devises as well as in the fundamental studies, aiming at the search of new microscopic
mechanisms of the ME coupling. Recently, the ME phenomenon is considered as a part of the more general paradigm called the `multiferroism', where
the appearance of spontaneous polarization is associated with some massive (and sometimes highly nontrivial) change of the
magnetic structure.\cite{MF_review}

  The canonical material exhibiting the ME effect is the rhombohedral Cr$_2$O$_3$, which was discussed by Dzyaloshinskii almost
six decades ago.\cite{Dzyaloshinskii} Particularly, the antiferromagnetic structure, realized in Cr$_2$O$_3$, is such that
the spacial inversion $\hat{I}$ enters the magnetic space group only in the combination with the time reversal $\hat{T}$.
Therefore, the application of either electric or magnetic field, which destroys $\hat{I}$ or $\hat{T}$, respectively, will destroy also
$\hat{I} \hat{T}$, thus, giving rise the electric polarization and the net magnetization. The symmetry properties of the
induced electric polarization depend on other symmetry operations, but the existence of $\hat{I} \hat{T}$ is
crucial for understanding the ME effect in Cr$_2$O$_3$.

  The ME effect of a similar origin has been discovered by Fischer \textit{et al.}\cite{Fischer} in 1972
in a family of $M_4$$A_2$O$_9$ materials,
where $M$ = Co or Mn and $A$ = Nb or Ta.
Like in Cr$_2$O$_3$, the magnetic structure of $M_4$$A_2$O$_9$ obeys the $\hat{I} \hat{T}$ symmetry, which can be
destroyed by either electric or magnetic field.
The interest to these materials has been recently revived in a series of papers which have addressed the details of the magnetic structure\cite{Bertaut,Schwarz,Cao,Khanh} and the ME coupling.\cite{Khanh,Kolodiazhnyi,Fang1,Fang2,Fang3}

For instance, with the availability of the single crystals it became possible to locate the easy axis for the magnetic moments\cite{Cao} and to solve the magnetic structure of Co$_4$Nb$_2$O$_9$ in the $C$2/$c^\prime$ magnetic space group\cite{Khanh} in contrast to the \emph{P}$\bar{3}$$^\prime$\emph{c}$^\prime$1 magnetic space group originally proposed by Bertaut et al.\cite{Bertaut} Furthermore, as reported by Khan et al.,\cite{Khanh} ME tensor in Co$_4$Nb$_2$O$_9$ shows several off-diagonal components which implies the existence of toroidal moment.

So far no evidence of type-II multiferroism has been reported for Co$_4$Ta$_2$O$_9$ and Co$_4$Nb$_2$O$_9$ and these compounds have been classified as linear magnetoelectrics where an applied electric, \emph{E} (magnetic, \emph{H}) field induces magnetization, $M$ (polarization, \emph{P}).\cite{Khanh,Fang1,Fang2}
In this contribution we report both experimental and first-principles analysis of the ME properties of Co$_4$Nb$_2$O$_9$ and Co$_4$Ta$_2$O$_9$.  Surprisingly, there is a significant difference in the magnitude of the ME effect in the Co$_4$Nb$_2$O$_9$ and Co$_4$Ta$_2$O$_9$ despite their identical crystal structure with the latter compound showing much weaker ME coupling at the N\'{e}el temperature, $T$$_{\mathrm{N}}$. In accordance with a recent report by Xie et al.\cite{Xie} we have detected a small anomaly in a zero-field dielectric permittivity at $T$ $\le$ $T$$_{\mathrm{N}}$ which confirms the magnetostriction effect in both compounds.
We will try to rationalize some of these data using results of first-principles electronic structure calculations.

  The rest of the article is organized as follows. In Sec.~\ref{sec:Experimental} we will present our experimental data for the magnetic
(Sec.~\ref{sec:Magnetic}) and magnetodielectric (Sec.~\ref{sec:Magnetodielectric}) properties of Co$_4$Nb$_2$O$_9$ and Co$_4$Ta$_2$O$_9$.
Results of theoretical calculations will be discussed in Sec.~\ref{sec:ElStr}.
Particularly, in Secs.~\ref{sec:ex} and \ref{sec:P} we will present our data, respectively, for the behavior of interatomic exchange interactions
and the polarization in the magnetic field. Finally, in Sec.~\ref{sec:conc} we discuss an overall picture emerging from the comparison of
experimental and theoretical data and draw our conclusions.

\section{\label{sec:Experimental} Experimental Results}
Samples were prepared from 99.99$\%$ pure Co$_3$O$_4$, Nb$_2$O$_5$ and 99.9$\%$ pure Ta$_2$O$_5$. The stoichiometric mixtures were treated at 900--1000 \C\ for 10 h. Dense ceramic bodies were obtained by 10 h sintering in air at 1050 and 1300 \C\ for Co$_4$Nb$_2$O$_9$ and Co$_4$Ta$_2$O$_9$, respectively. Phase purity was confirmed by powder X-ray diffraction (MiniFlex600 diffractometer with Cu \emph{K}$_{\alpha}$ x-ray source, Rigaku, Japan). Lattice parameters were obtained from Rietveld refinement of the X-ray data using JANA2006.\cite{JANA2006} Magnetic susceptibility in the 2--100 K range was measured using superconducting quantum interference devise (Magnetic Property Measurement System, Quantum Design). For dielectric measurements, Au electrodes were spattered on the 1 mm thick disc-shaped ceramic samples. Dielectric properties in the 10 Hz -- 1 MHz frequency range were measured
with Novocontrol Alpha Impedance analyzer in the temperature interval of 2--40 K and magnetic fields up to 9 T. For temperature and magnetic field dependence of the dielectric properties we have used a home-made sample cell coupled with commercial cryostat equipped with superconducting magnet (Physical Property Measurement System, Quantum Design).

\subsection{\label{sec:Magnetic} Magnetic properties}
\begin{figure}[tbp]
\begin{center}
\scalebox{0.4}{\includegraphics[0,0][523,488]{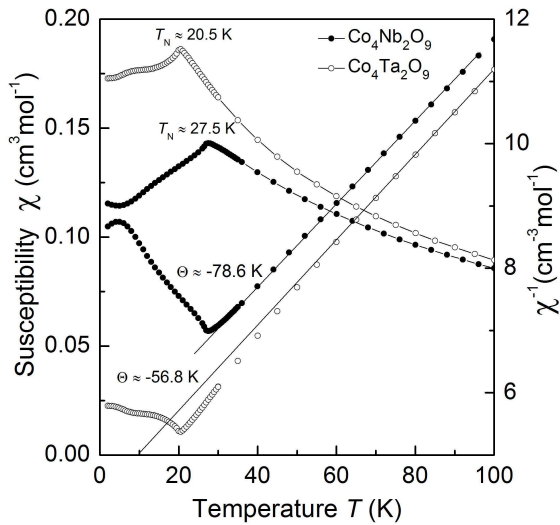}}
\end{center}
\caption{
Temperature dependence of magnetic susceptibility of Co$_4$Nb$_2$O$_9$ and Co$_4$Ta$_2$O$_9$.}
\label{fig.magsus}
\end{figure}
Temperature dependence of magnetic susceptibility, $\chi$, determined here as a ratio of magnetic moment, $M$, over magnetic field, $H$ is shown in Fig.~\ref{fig.magsus} together with inverse magnetic susceptibility for Co$_4$Nb$_2$O$_9$ and Co$_4$Ta$_2$O$_9$ powder samples measured at $H$ = 50 Oe. The $\chi$($T$) shows a cusp at $T$$_{\mathrm{N}}$ $\approx$ 27.5 and 20.5 K for Co$_4$Nb$_2$O$_9$ and Co$_4$Ta$_2$O$_9$, respectively, in agreement with literature data.\cite{Khanh,Fang1,Fang2} Below $T$$_{\mathrm{N}}$ the $\chi$($T$) decreases followed by a gradual increase at the lowest temperatures. This can be understood as an effect of anisotropic $\chi$($T$) behaviour as explained in detail in Refs.~\onlinecite{Cao,Khanh}.
The fit of inverse magnetic susceptibility to the Curie-Weiss law $\chi$($T$) = C/($T$ - $\Theta$), gives Weiss temperature of $\Theta$ $\approx$ $-$$78.6$ and $-$$56.8$ K and effective magnetic moments of $\mu_{eff}$ $\approx$ 5.56 and 5.36 $\mu_B$ for Co$_4$Nb$_2$O$_9$ and Co$_4$Ta$_2$O$_9$, respectively. The $\mu_{eff}$ of the title compounds are significantly larger than the spin-only effective magnetic moment of Co$^{2+}$ $S$ = 3/2 $\mu_{eff}$ = 3.87 $\mu_B$ and indicate that orbital angular momentum is not quenched in the degenerate ground state, so there is significant contribution from the orbital angular magnetic moment.

\begin{figure}[tbp]
\begin{center}
\scalebox{0.35}{\includegraphics[0,0][478,492]{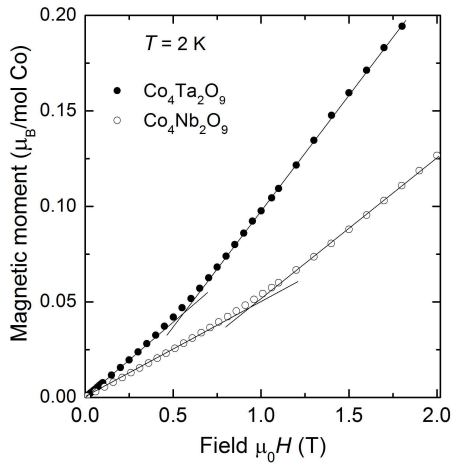}}
\end{center}
\caption{
Magnetic field dependence of magnetic moment of Co$^{2+}$ ion in Co$_4$Nb$_2$O$_9$ and Co$_4$Ta$_2$O$_9$ measured at 2 K.}
\label{fig.spinflop}
\end{figure}

\begin{figure}[tbp]
\begin{center}
\includegraphics[width=10cm]{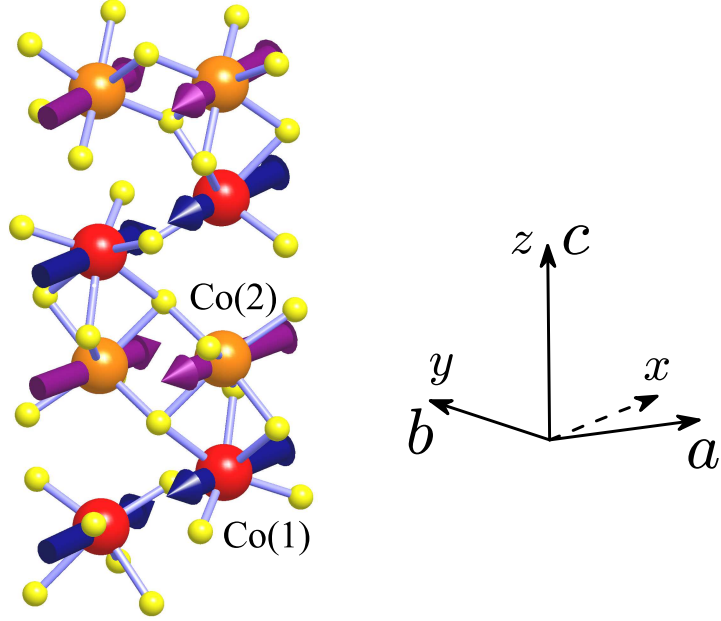}
\end{center}
\caption{(Color online)
Fragment of the crystal and magnetic structure of Co$_4$Nb$_2$O$_9$ in the ground state.
The relative orientation of hexagonal ($a$, $b$, and $c$) and cartesian ($x$, $y$, and $z$)
axes is explained on the right.}
\label{fig.str}
\end{figure}

Fig.~\ref{fig.spinflop} shows magnetic field dependence of Co$^{2+}$ magnetic moment of the title compounds measured at 2 K. The change in the slope of the magnetization attributed to the spin-flop phase transition is detected at $\mu_0$$H_c$ $\sim$ 0.9 T and $\sim$ 0.5 T for Co$_4$Nb$_2$O$_9$ and Co$_4$Ta$_2$O$_9$, respectively. Clearly, Co$_4$Ta$_2$O$_9$ is magnetically softer than the Co$_4$Nb$_2$O$_9$. Please note that our data on $H_c$ refer to the `average' critical field for polycrystalline materials. They are significantly larger than the $H_c$ $\sim$ 0.2 T measured along the easy plane $\emph{\textbf{H}}//x$ of the Co$_4$Nb$_2$O$_9$ single crystal. The orientation of the crystal is explained in Fig.~\ref{fig.str}.
Note that in our notations $x$, $y$, and $z$ correspond to, respectively,
$[1\bar{1}0]$, $[\bar{1}\bar{1}0]$, and $[001]$, in the notations of Ref.~\onlinecite{Khanh}.

\subsection{\label{sec:Magnetodielectric} Magnetodielectric properties}
\begin{figure}[tbp]
\begin{center}
\scalebox{0.5}{\includegraphics[0,0][359,323]{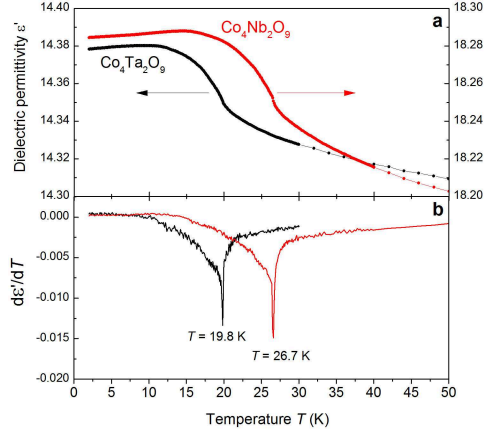}}
\end{center}
\caption{
Zero field dielectric permittivity (a) and its temperature derivative (b) for Co$_4$Nb$_2$O$_9$ and Co$_4$Ta$_2$O$_9$ measured at 250 kHz.}
\label{fig.diel0T}
\end{figure}

Room temperature bulk resistivities of the title compounds are 3.3 $\times$ 10$^5$ and 4.4 $\times$ 10$^4$ $\Omega$~cm for Co$_4$Nb$_2$O$_9$ and Co$_4$Ta$_2$O$_9$, respectively. Therefore, the dielectric permittivity, $\varepsilon$$^{\prime}$, above 100 K is dominated by the Maxwell-Wagner effect due to the electrode and grain boundary polarization. Below $\sim$ 50 K the free carriers are frozen in and dielectric properties are determined by the bulk of the crystals. The low-temperature dependence of the $\varepsilon$$^{\prime}$ and its first derivative measured at 0 T are shown in Fig.~\ref{fig.diel0T}.
An anomaly in the $\varepsilon$$^{\prime}$($T$) dependence for both Co$_4$Nb$_2$O$_9$ and Co$_4$Ta$_2$O$_9$ is attributed to the antiferromagnetic (AFM) phase transition. This indicates a small but finite coupling of the spin ordering to the dielectric response in zero magnetic field in the title compounds. Note that earlier studies\cite{Fang1,Fang2,Kolodiazhnyi} could not detect the ME effect in Co$_4$Nb$_2$O$_9$ and Co$_4$Ta$_2$O$_9$ at zero magnetic field. Very recently magnetoelastic coupling was proposed in Co$_4$Nb$_2$O$_9$ based on the anomaly in the lattice strain below $T$$_{\mathrm{N}}$.\cite{Xie} Magnetoelastic contraction of the lattice can be partially responsible for a notable upturn in the $\varepsilon$$^{\prime}$($T$) below $T$$_{\mathrm{N}}$ in both compounds shown in Fig.~\ref{fig.diel0T}a in which case one would expect slight re-normalization of the phonon frequencies due to the spin-phonon coupling. This conclusion is supported by the absence of the frequency dispersion of the dielectric permittivity below N\'{e}el temperature reported in literature\cite{Xie} and also confirmed in this work.


\begin{figure}[tbp]
\begin{center}
\scalebox{0.5}{\includegraphics[0,0][514,360]{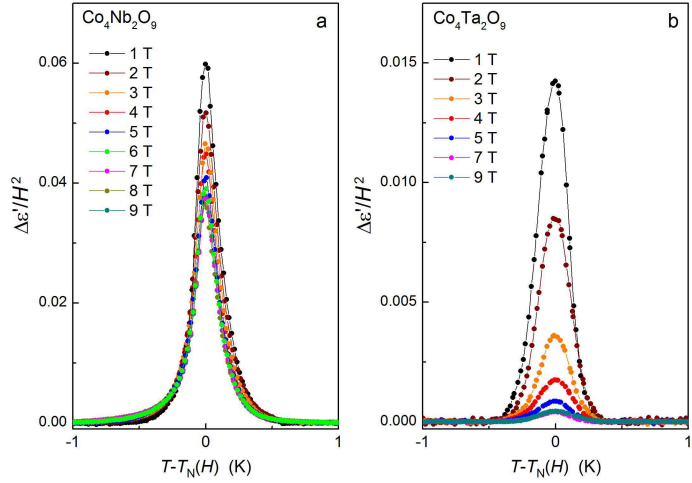}}
\end{center}
\caption{
Scaled dielectric permittivity with subtracted zero-field background, ($\varepsilon$$^{\prime}$($H$) -- $\varepsilon$$^{\prime}$($H$ = 0))/$H^2$, versus $T$ -- $T$$_{\mathrm{N}}$. Dielectric permittivity was measured at 1 kHz.}
\label{fig.EoverH2}
\end{figure}

\begin{figure}[tbp]
\begin{center}
\scalebox{0.5}{\includegraphics[0,0][447,322]{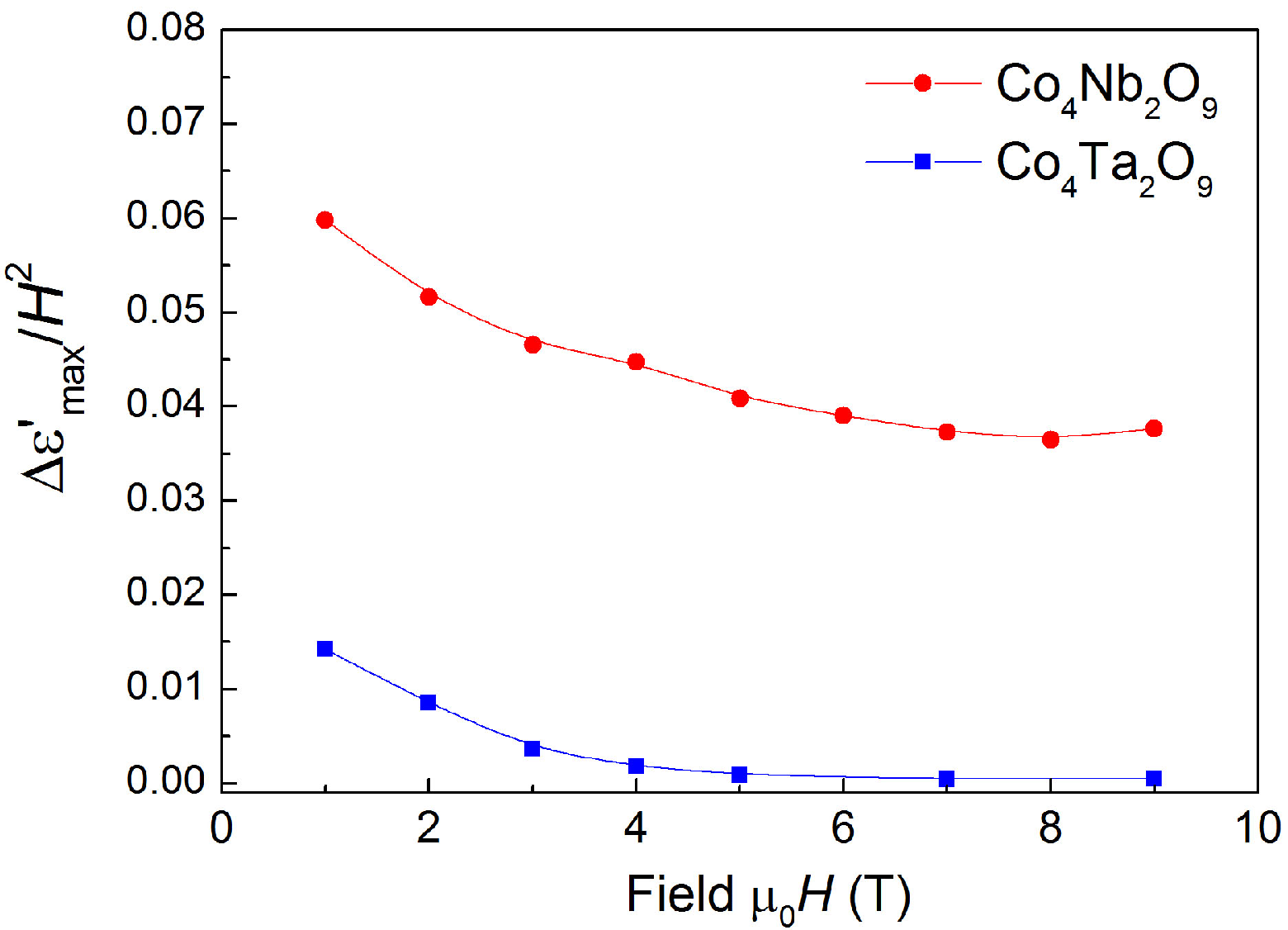}}
\end{center}
\caption{
Magnetic field dependence of the maximum of the scaled dielectric permittivity at $T$$_{\mathrm{N}}$.}
\label{fig.EmaxoverH2}
\end{figure}

In contrast to our expectations, the two title compounds show quantitatively different magnetocapacitance effect. The effect of magnetic field on the scaled dielectric permittivity with subtracted zero-field background, $\Delta$$\varepsilon$$^{\prime}$/$H^2$ = ($\varepsilon$$^{\prime}$($H$) -- $\varepsilon$$^{\prime}$($H$ = 0))/$H^2$, versus $T$ -- $T$$_{\mathrm{N}}$ is shown in Figure \ref{fig.EoverH2}. According to our data, at 1 T the $\Delta$$\varepsilon$$^{\prime}$/$H^2$ value at $T$$_{\mathrm{N}}$ is $\sim$ 4 times higher for Co$_4$Nb$_2$O$_9$ than for Co$_4$Ta$_2$O$_9$. The difference increases to $\sim$ 88 at magnetic field of 7 T. Magnetic-field induced spontaneous polarization, $P$, for polycrystalline Co$_4$Nb$_2$O$_9$ and  Co$_4$Ta$_2$O$_9$ reported in the literature\cite{Fang1,Fang2} equal to $P$ $\approx$ 120 $\mu$C/m$^2$ and 44 $\mu$C/m$^2$ at 7 T, respectively. Because the dielectric susceptibility at the phase transition is equal to the induced polarization response to a small electric stimuli, the factor of 2.7 difference in $P$ is in reasonable agreement with a factor of 4 difference in mangetocapacitance at low magnetic field.

According to the model developed for linear magnetoelectric, e.g., MnTiO$_3$,\cite{Mufti} the $\Delta$$\varepsilon$$^{\prime}$ is expected to scale linearly with $H^2$ which should yield a field-independent value of the $\Delta$$\varepsilon$$^{\prime}$/$H^2$ maximum. As evidenced from the Fig.~\ref{fig.EmaxoverH2} this is not exactly the case for Co$_4$Nb$_2$O$_9$ and Co$_4$Ta$_2$O$_9$. The $\Delta$$\varepsilon$$^{\prime}$/$H^2$ shows a notable decrease with increasing magnetic field before it levels off at $\mu_0$$H$ $\ge$ 7 and 5 T for Co$_4$Nb$_2$O$_9$ and Co$_4$Ta$_2$O$_9$, respectively. The saturation of the $\Delta$$\varepsilon$$^{\prime}$/$H^2$ maximum is difficult to attribute to the spin flop phase transition because the $\Delta$$\varepsilon$$^{\prime}$/$H^2$ saturation occurs at the fields significantly higher than the spin flop critical fields $\mu_0$$H_c$ of less than 1 T for both compounds.
To gain further insight into the origin of the ME effect below we report the results of the first-principles calculations.

\section{\label{sec:ElStr} Theoretical Calculations}

\subsection{\label{sec:Structure} Main details of crystal and electronic structure}
\begin{figure}[tbp]
\begin{center}
\includegraphics[width=7cm]{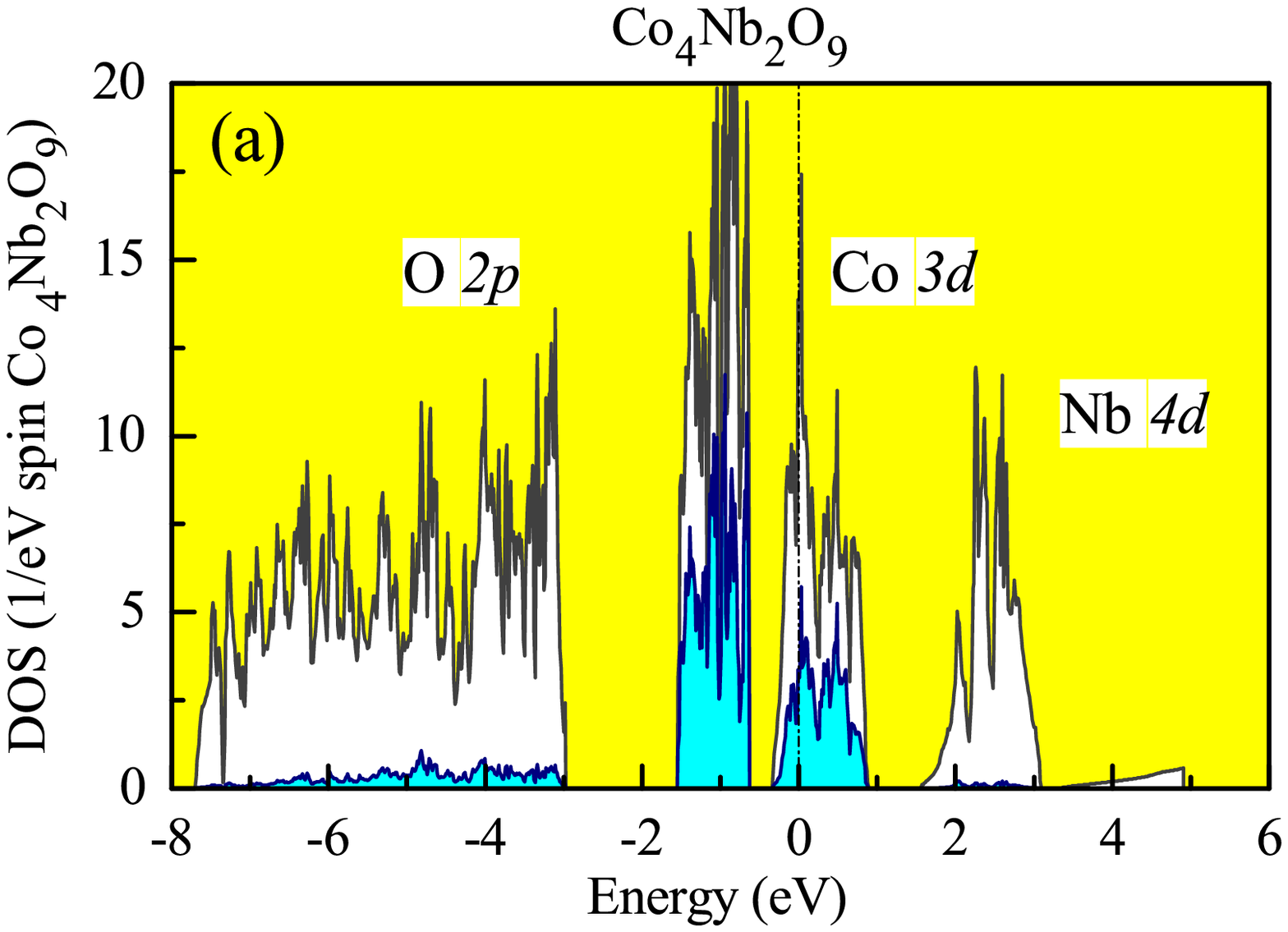} \qquad
\includegraphics[width=7cm]{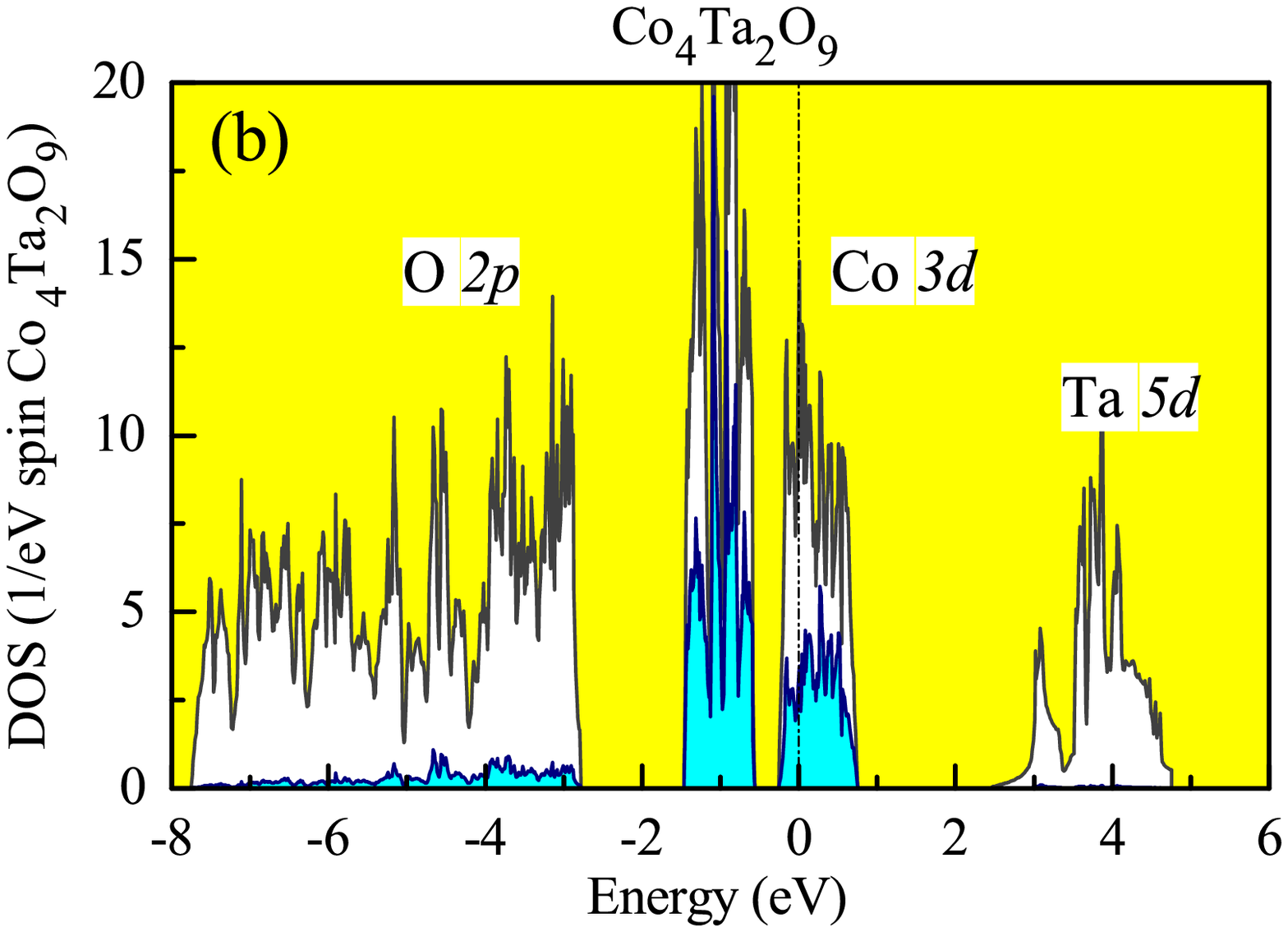}
\end{center}
\caption{(Color online)
Total and partial densities of states of Co$_4$Nb$_2$O$_9$ and Co$_4$Ta$_2$O$_9$ in the local density approximation.
The shaded light (blue) area shows contributions of the Co$3d$ states.
Positions of the main bands are indicated by symbols. The Fermi level is at zero energy (shown by dot-dashed line).}
\label{fig.DOS}
\end{figure}

Co$_4$Nb$_2$O$_9$ and Co$_4$Ta$_2$O$_9$ crystallize in the centrosymmetric trigonal $P\overline{3}c1$ structure (No. 165).
The fragment of this structure is shown in Fig.~\ref{fig.str}.
There are two inequivalent types of Co atoms, alternating along the $z$ axis and forming the (distorted)
honeycomb layers in the $xy$-plane. In our electronic structure calculations, we use the experimental
room-temperature atomic positions and lattice parameters, reported in Refs.~\onlinecite{Co4Nb2O9str} and \onlinecite{Co4Ta2O9str}
for Co$_4$Nb$_2$O$_9$ and Co$_4$Ta$_2$O$_9$, respectively. The total and partial densities of states, obtained
in the local-density approximation (LDA), are explained in Fig.~\ref{fig.DOS}.
The states located near the Fermi level are the Co$3d$ bands, which in the octahedral CoO$_6$ environment
are split into lower-energy $t_{2g}$ and higher-energy $e_g$ bands. These bands are mainly responsible for the
magnetic properties of Co$_4$Nb$_2$O$_9$ and Co$_4$Ta$_2$O$_9$.
Regarding the LDA band structure, there are two main differences between Co$_4$Nb$_2$O$_9$ and Co$_4$Ta$_2$O$_9$.
The Ta$5d$ states are considerably more extended in comparison with the Nb$4d$ ones, and, therefore,
much stronger hybridize with the O$2p$ states.
This explains the additional upward shift of the antibonding Ta$5d$ band,
arising from the O$2p$-Ta$5d$ hybridization. This in turn reduces the Co$3d$-Ta$5d$ hybridization
and slightly decreases the width of the Co$3d$ bands in Co$_4$Ta$_2$O$_9$. In this sense, the Co$3d$ states in
Co$_4$Ta$_2$O$_9$ are slightly more ``localized'' in comparison with Co$_4$Nb$_2$O$_9$.

\subsection{\label{sec:elmodel} The effective low-energy electron model}
In this section, we briefly remind the reader the main steps of the construction and solution of the
effective low-energy model, which is used for the analysis of electronic and magnetic properties of Co$_4$Nb$_2$O$_9$ and Co$_4$Ta$_2$O$_9$.

  The first step of our approach
is the construction of the effective Hubbard-type model for the magnetically active Co$3d$ bands:
\begin{equation}
\hat{\cal{H}}  =  \sum_{ij} \sum_{\sigma \sigma'} \sum_{ab}
\left( t^{ij}_{ab}\delta_{\sigma \sigma'} + \Delta t^{i \sigma \sigma'}_{ab}\delta_{ij} \right)
\hat{c}^\dagger_{i a \sigma}
\hat{c}^{\phantom{\dagger}}_{j b \sigma'} +
  \frac{1}{2}
\sum_{i}  \sum_{\sigma \sigma'} \sum_{abcd} U^i_{abcd}
\hat{c}^\dagger_{i a \sigma} \hat{c}^\dagger_{i c \sigma'}
\hat{c}^{\phantom{\dagger}}_{i b \sigma}
\hat{c}^{\phantom{\dagger}}_{i d \sigma'},
\label{eqn.ManyBodyH}
\end{equation}
starting from the electronic band structure in LDA. The model itself is formulated in the basis of
Wannier functions, which were obtained using the projector-operator technique (Refs.~\onlinecite{review2008,WannierRevModPhys})
and the orthonormal linear muffin-tin orbitals (LMTO's, Ref.~\onlinecite{LMTO}) as the trial wave functions.
$\sigma (\sigma')$$=$ $\uparrow$ or $\downarrow$ in (\ref{eqn.ManyBodyH}) are the spin indices, while $a$, $b$, $c$, and $d$ label
five $3d$ orbitals.
The parameters of the one-electron part,
$\hat{t} = [t^{ij}_{ab}]$, are defined as the matrix elements of the LDA Hamiltonian in the Wannier basis.\cite{review2008}
$\Delta \hat{t} = [\Delta t^{i \sigma \sigma'}_{ab}]$ is the matrix of spin-orbit (SO) interaction, also in the Wannier basis.
The parameters of screened on-site Coulomb interactions, $\hat{U} = [U^i_{abcd}]$, are calculated in the framework of
constrained random-phase approximation (RPA),\cite{Ferdi04} using the simplified procedure, which was explained in Ref.~\onlinecite{review2008}.

\begin{figure}[tbp]
\begin{center}
\includegraphics[width=10cm]{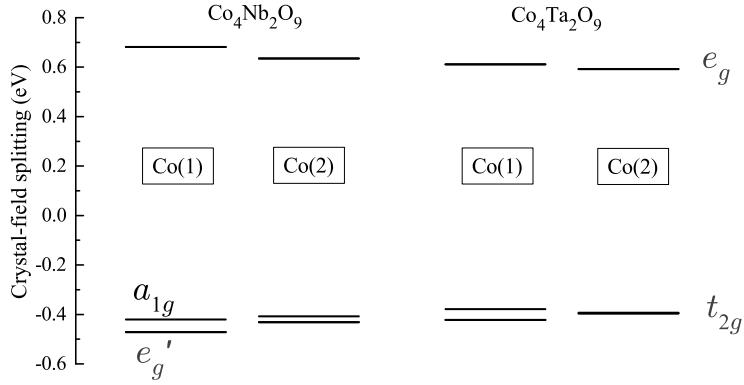}
\end{center}
\caption{
Scheme of the crystal-field splitting for two inequivalent Co sites
in Co$_4$Nb$_2$O$_9$ and Co$_4$Ta$_2$O$_9$.}
\label{fig.CF}
\end{figure}

  The crystal-field splitting, obtained from the diagonalization of the site-diagonal part of $\hat{t}$,
is shown in Fig.~\ref{fig.CF}.
One can clearly see that the main effect is the $t_{2g}$-$e_g$ splitting in the octahedral CoO$_6$ environment.
Other splittings are considerably smaller. For instance, three $t_{2g}$ levels are split into lower-energy doublet ($e_g'$) and
higher-energy singlet ($a_{1g}$), being consistent with the $d^7$ configuration of Co$^{2+}$, where two minority-spin electrons
are accommodated in the lower-energy doublet. The largest splitting of the $t_{2g}$ levels is about 50 meV,
which is comparable with the strength of the SO coupling $\xi$ (about 75 meV).
Thus, one can expect the existence of unquenched orbital magnetization, which is supported by the
experimental susceptibility data.

  Each $5$$\times$$5$$\times$$5$$\times5$ matrix $\hat{U} = [U^i_{abcd}]$ can be parameterized in terms of three parameters:
the Coulomb repulsion $U=F^0$, the
intraatomic exchange interaction $J = (F^2$$+$$F^4)/14$, and the `nonsphericity'
$B = (9F^2$$-$$5F^4)/441$, where $F^0$, $F^2$, and $F^4$ are
the screened radial Slater's integrals. The results of such parametrization are shown in Table~\ref{tab:scrint}.
\begin{table}[h!]
\caption{Parameters of screened Coulomb interaction ($U$), exchange interaction ($J$) and nonsphericity ($B$)
for the inequivalent Co sites in Co$_4$Nb$_2$O$_9$ and Co$_4$Ta$_2$O$_9$ (in eV).}
\label{tab:scrint}
\begin{ruledtabular}
\begin{tabular}{lcccc}
 & \multicolumn{2}{c}{Co$_4$Nb$_2$O$_9$} & \multicolumn{2}{c}{Co$_4$Ta$_2$O$_9$} \\
 \cline{2-3} \cline{4-5}
                & Co(1)  & Co(2)
                & Co(1)  & Co(2) \\
\hline
$U$ & $2.95$ & $3.00$  & $3.00$ & $3.29$  \\
$J$ & $0.94$ & $0.93$  & $0.94$ & $0.95$  \\
$B$ & $0.10$ & $0.10$  & $0.10$ & $0.10$  \\
\end{tabular}
\end{ruledtabular}
\end{table}
The parameters $U$ are generally larger in Co$_4$Ta$_2$O$_9$. This is due to the additional energy separation between the
Co$3d$ and Ta$5d$ bands (see Fig.~\ref{fig.DOS}),
which results in less efficient screening of the Coulomb interactions in the Co$3d$ band by the
Ta$5d$ band.\cite{review2008} Moreover, due to different crystallographic environment, the Coulomb $U$ is different for the Co sites 1 and 2,
and this difference is substantially larger in Co$_4$Ta$_2$O$_9$.

  Other parameters of the model Hamiltonian can be found elsewhere.\cite{request}

\subsection{\label{sec:ex} Magnetic interactions and magnetic ground state}
After the construction, the model was solved in the mean-field Hartree-Fock approximation.\cite{review2008}
Then, the isotropic exchange interactions can be evaluated by considering the
infinitesimal rotations of spins and mapping corresponding energy changes onto the
spin Heisenberg model $E_{\rm H} = -\sum_{i>j} J_{j} \boldsymbol{e}_i \cdot \boldsymbol{e}_{i+j}$, where $\boldsymbol{e}_i$
denotes the \textit{direction of spin} at the site $i$.\cite{JHeisenberg}

\begin{figure}[tbp]
\begin{center}
\includegraphics[width=10cm]{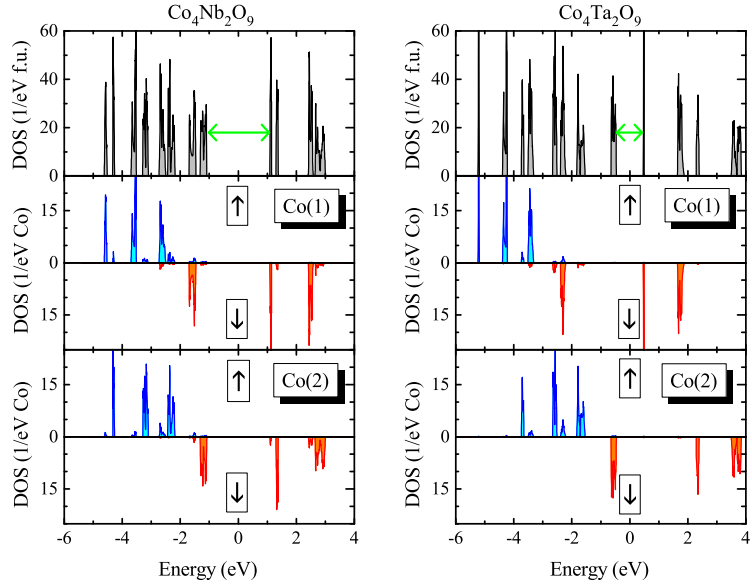}
\end{center}
\caption{(Color online)
Total and partial densities of states for the two inequivalent Co-sites in Co$_4$Nb$_2$O$_9$ and Co$_4$Ta$_2$O$_9$,
as obtained in the model Hartree-Fock calculations for the antiferromagnetic ground state.
The green arrows show the band gap, which is formed between states of the inequivalent Co-atoms.
The Fermi level is at zero energy.}
\label{fig.DOSHF}
\end{figure}

  We have found that the lowest energy
corresponds to the AFM ground state,
where all the spins are coupled ferromagnetically along the $z$ axis
and antiferromagnetically in the $xy$ plane (see Fig.~\ref{fig.str}),
being in total agreement with the experimental data.\cite{Fang1,Khanh}
The corresponding electronic structure is shown in Fig.~\ref{fig.DOSHF}.
The $3d$ states in Co$_4$Nb$_2$O$_9$ and Co$_4$Ta$_2$O$_9$ are indeed well localized:
the atomic levels are split by large crystal field and on-site Coulomb interactions. The interatomic transfer integrals
are considerably weaker and lead to the formation of narrow bands
around the atomic levels. The main difference between
Co$_4$Nb$_2$O$_9$ and Co$_4$Ta$_2$O$_9$ is the additional upward shift of the Co(2) states in the latter compound
due to larger Coulomb repulsion (see Table~\ref{tab:scrint}).
In particular, it reduces the band gap in Co$_4$Ta$_2$O$_9$, which is formed between minority-spin states of
the atoms Co(2) and Co(1).

\begin{figure}[tbp]
\begin{center}
\includegraphics[width=10cm]{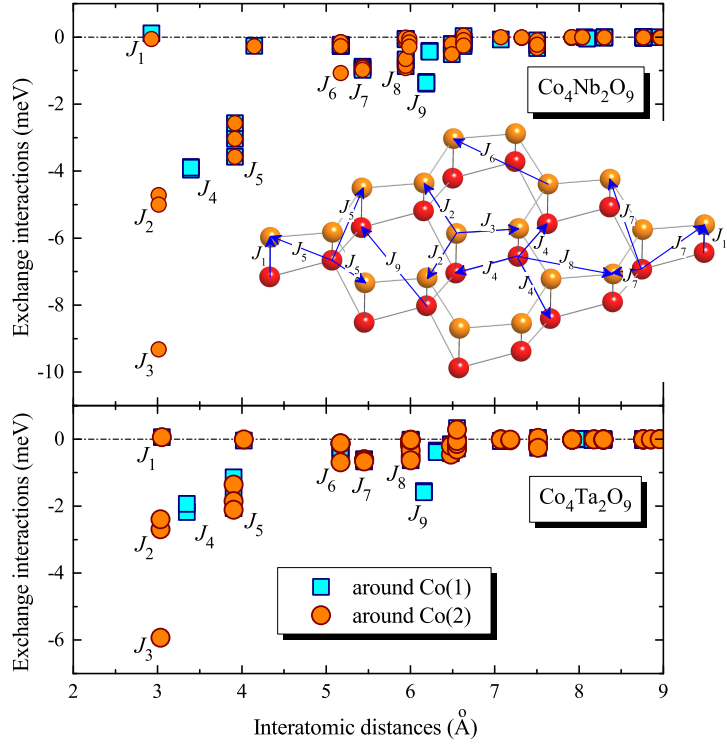}
\end{center}
\caption{(Color online)
Distance dependance of interatomic exchange interactions in Co$_4$Nb$_2$O$_9$ and Co$_4$Ta$_2$O$_9$
around two inequivalent types of Co atoms. The main exchange interactions are explained in the inset,
where the Co-atoms of the first and second types are shown by darker (red) and lighter (orange)
spheres, respectively.}
\label{fig.J}
\end{figure}

The obtained type of the magnetic ground state
can be easily understood by considering the behavior of interatomic exchange interactions (Fig.~\ref{fig.J}).
One can see that the main interactions are AFM. Partly, this is an artifact of our model analysis, because it does not take
into account the polarization of the O$2p$ states, which gives rise to important ferromagnetic (FM) contributions
to the exchange coupling.
For instance, the N\'eel temperature, estimated using the calculated parameters is about $170$ K and $100$ K
for Co$_4$Nb$_2$O$_9$ and Co$_4$Ta$_2$O$_9$, respectively. It is strongly overestimated in comparison with the
experimental data ($27.5$ K and $20.5$ K for Co$_4$Nb$_2$O$_9$ and Co$_4$Ta$_2$O$_9$, respectively).
This overestimation is partly caused by the mean-field approximation. However, there is also an intrinsic error, inherent to the
low-energy model itself, which does not take into account the FM contributions
caused by the polarization of the O$2p$ states.
Similar overestimation has been found for the Weiss temperature
($-$$255$ K and $-$$155$ K, for Co$_4$Nb$_2$O$_9$ and Co$_4$Ta$_2$O$_9$, respectively). We would like to note that this problem is not new
and was also encountered in other systems, which are close to the charge transfer regime, and
where the oxygen states play a more important role and, as a rule, should be explicitly treated in the
model analysis.\cite{Ba2CoGe2O7,CrO2}
The correct quantitative description is possible by considering the direct exchange interactions and the
magnetic polarization of the O$2p$ band.\cite{CrO2}
Nevertheless, the magnetic interactions, obtained in the present low-energy model,
are well consistent with the
observed AFM ground state: the main interactions, stabilizing the AFM alignment in the $xy$ plane, are
$J_2$, $J_3$, and $J_4$. In the combination with the AFM interaction $J_5$, they also stabilize the FM alignment along $z$.
The AFM interactions are systematically weaker in Co$_4$Ta$_2$O$_9$ (and, therefore, magnetically this system is
expected to be softer). This is consistent with somewhat narrower Co$3d$ bandwidth (Fig.~\ref{fig.DOS}),
larger values of $U$, which weakens the superexchange interactions, and also smaller bandgap between the minority-spin
states, which enlarges the FM contributions to the superexchange coupling.\cite{KugelKhomskii}

  After turning on the SO coupling, the magnetic moments become aligned mainly in the $xy$ plane. In this case,
we have found two nearly degenerate solutions with the magnetic moments being mainly parallel
to either $x$ or $y$ axes and obeying the following symmetries (the magnetic space groups), respectively,
$\boldsymbol{G}_1 = \{\hat{E}, \hat{T}\hat{I}, \hat{T}\hat{m}_y, \hat{C}^2_y \}$
(or $C2/c'$ in the notations of Ref.~\onlinecite{Khanh})
and
$\boldsymbol{G}_2 = \{\hat{E}, \hat{T}\hat{I}, \hat{m}_y, \hat{T}\hat{C}^2_y \}$, where
$\hat{E}$ is the unity operation, $\hat{I}$ is the inversion, $\hat{m}_y$ and $\hat{C}^2_y$ are, respectively,
the mirror reflection and the $180^\circ$ rotation about the $y$ axis, combined with the half of the
hexagonal translation, $\boldsymbol{c}/2$, and $\hat{T}$ is the time inversion operation.
The threefold rotation about the $z$ ($c$) axis, which is the symmetry operation of the parent
space group $P\overline{3}c1$, is forbidden by the magnetic alignment in the $xy$ plane.
The first such solution is illustrated in Fig.~\ref{fig.str}.
Both in $\boldsymbol{G}_1$ and $\boldsymbol{G}_2$, the magnetic moments exhibit the AFM canting out of the main
(either $x$ or $y$) axis
due to the joint effect of the single-ion anisotropy and Dzyaloshinskii-Moriya interactions.
The canting angle at the site Co(1) and Co(2) in Co$_4$Nb$_2$O$_9$ (Co$_4$Ta$_2$O$_9$)
is about $2^{\circ}$ ($1^{\circ}$) and $6^{\circ}$ ($7^{\circ}$), respectively.
This canting is considerably smaller than the experimental one,\cite{Khanh}
due to the overestimation of isotropic exchange interactions in the
low-energy model, which makes these magnetic materials substantially harder than
in the experiment.
Moreover, we have found an appreciable orbital contribution, which constitutes about 17-20\%
of the total magnetization at the Co site in the ground state.

\subsection{\label{sec:P} Magnetic-field dependence of electric polarization}
The electronic polarization in the external magnetic field can be
computed in the reciprocal space,
using the formula of King-Smith and Vanderbilt:\cite{FE_theory}
\begin{equation}
\textbf{P} = - \frac{ie}{(2 \pi)^3} \sum_{n}
\int_{BZ} d \textbf{k} \,  \langle u_{n \textbf{k}} | \partial_{\textbf{k}} u_{n \textbf{k}} \rangle ,
\label{eqn:PKSV}
\end{equation}
where $u_{n \textbf{k}}(\textbf{r}) = e^{-i \textbf{kr}} \psi_{n \textbf{k}}(\textbf{r})$ is the cell-periodic eigenstate
of the model Hamiltonian $H_{\textbf{k}} = e^{-i \textbf{kr}} H e^{i \textbf{kr}}$,
which in our case is treated in the Hartree-Fock (HF) approximation,
the summation runs over the
occupied bands ($n$), the $\textbf{k}$-space integration goes over the first
Brillouin zone (BZ), and $-$$e$ ($e > 0$) is the electron charge. Since the Co$3d$ states in Co$_4$Nb$_2$O$_9$ and Co$_4$Ta$_2$O$_9$
are well localized (see Fig.~\ref{fig.DOSHF}), the analysis can be also performed in the real space,
starting from the limit of atoms states and using the perturbation theory expansion with
respect to the transfer integrals.\cite{PRB14}

\begin{figure}[tbp]
\begin{center}
\includegraphics[width=10cm]{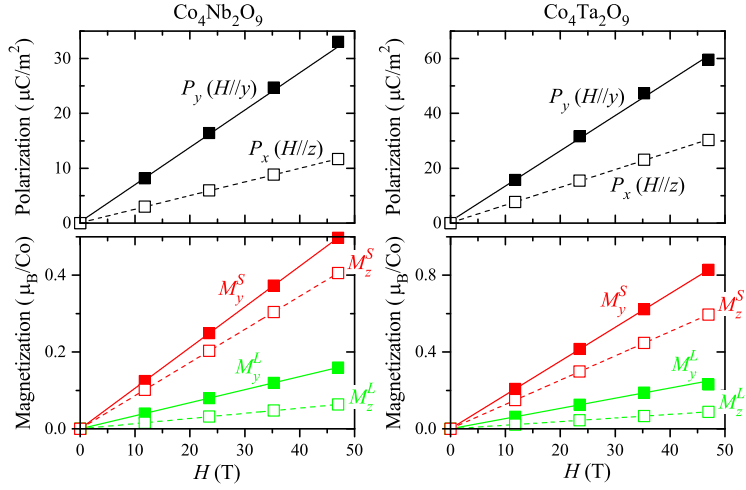}
\end{center}
\caption{(Color online)
Nonvanishing components of electric polarization ($P$) and the net spin ($M^S$)
and orbital ($M^L$) magnetization as obtained
in the model Hartree-Fock calculations for the $\boldsymbol{G}_1$ state
in the magnetic field being parallel to either
$y$ or $z$ axes.}
\label{fig.PG1NbTa}
\end{figure}

\begin{figure}[tbp]
\begin{center}
\includegraphics[width=10cm]{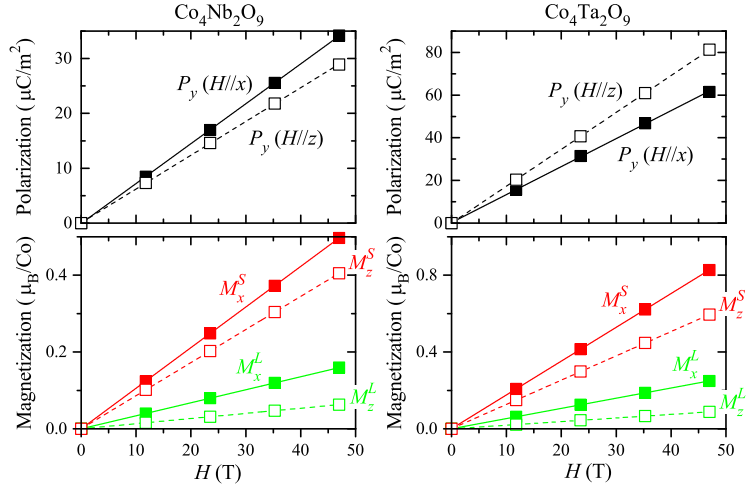}
\end{center}
\caption{(Color online)
Nonvanishing components of electric polarization ($P$) and the net spin ($M^S$)
and orbital ($M^L$) magnetization as obtained
in the model Hartree-Fock calculations for the $\boldsymbol{G}_2$ state
in the magnetic field being parallel to either
$x$ or $z$ axes.}
\label{fig.PG2NbTa}
\end{figure}

  Since $\hat{T}\hat{I}$ is one of the symmetry operations in $\boldsymbol{G}_1$ and $\boldsymbol{G}_2$,
these states develop neither spontaneous polarization $\textbf{P}$ nor the net magnetization $\textbf{M}$.
However, both polarization and magnetization can be induced by the magnetic field,
which breaks $\hat{T}\hat{I}$.\cite{Dzyaloshinskii}
When applied along the $x$, $y$, or $z$ axis in the $\boldsymbol{G}_1$ state it, respectively,
causes the spin-flop transition to the $\boldsymbol{G}_2$ state, breaks the $\hat{T}\hat{m}_y$ symmetry
(while the symmetry operation $\hat{C}^2_y$ remains) and induces the polarization parallel to the $y$ axis,
and breaks the $\hat{C}^2_y$ symmetry (while $\hat{T}\hat{m}_y$ remains) and induces the polarization in the $zx$ plane.
Moreover, we have found that the $z$ component of the electric polarization is negligibly small,
which is consistent with the experimental data.\cite{Khanh}
The results of numerical simulations are presented in Fig.~\ref{fig.PG1NbTa}.
In the $\boldsymbol{G}_2$ state, the magnetic field applied along either $x$ or $z$ axes, breaks the $\hat{m}_y$ symmetry
(while the symmetry operation $\hat{T}\hat{C}^2_y$ remains) and induces the electric polarization
parallel to the $y$ axis (Fig.~\ref{fig.PG2NbTa}).
The magnetic field along the $y$ axis causes the spin-flop transition to the $\boldsymbol{G}_1$ state.
The toroidal moment $\textbf{T} \sim (\textbf{P} \times \textbf{M})$ is expected in the $\boldsymbol{G}_1$ state
when $\emph{\textbf{H}} // z$ and in the $\boldsymbol{G}_2$ state when $\emph{\textbf{H}} // x$ or $z$.\cite{Khanh}

\begin{figure}[tbp]
\begin{center}
\includegraphics[width=10cm]{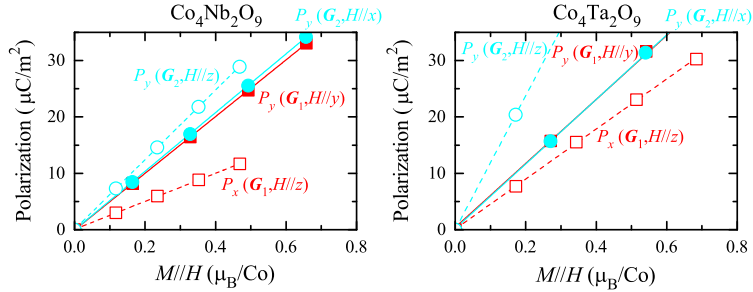}
\end{center}
\caption{
Electric polarization versus total (spin plus orbital) net magnetization
as obtained
in the model Hartree-Fock calculations for
different directions of the magnetic field in the states $\boldsymbol{G}_1$ and $\boldsymbol{G}_2$.}
\label{fig.PvM}
\end{figure}

  The calculated polarization is generally larger in Co$_4$Ta$_2$O$_9$. This is because of two factors.
On the one hand, Co$_4$Ta$_2$O$_9$ is magnetically softer than Co$_4$Nb$_2$O$_9$ and, therefore,
the magnetic structure of Co$_4$Ta$_2$O$_9$ can be easier deformed by the magnetic field. This effect alone
nicely explains the behavior of the electric polarization, when the magnetic field is applied in the
$xy$ plane: the dependence of $P_y$ on the total (spin plus orbital) net magnetization is practically identical
for Co$_4$Nb$_2$O$_9$ and Co$_4$Ta$_2$O$_9$ (see Fig.~\ref{fig.PvM}).
Therefore, for a given magnetic field $H$, $P_y$ is larger in Co$_4$Ta$_2$O$_9$ only because this field
induces larger net magnetization. Another factor, which further increases
$P_x$ and $P_y$ (for $\boldsymbol{G}_1$ and $\boldsymbol{G}_2$, respectively)
in the case of Co$_4$Ta$_2$O$_9$, is smaller band gap (see Fig.~\ref{fig.DOSHF}).
Note that in the atomic limit the electric polarization is inversely proportional to the
splitting between the occupied and unoccupied atomic levels.\cite{PRB14}
Since the band gap is formed between the states of the atoms Co(1) and Co(2), alternating along the $z$ axis (see Fig.~\ref{fig.str}),
this effect will be more important for $\emph{\textbf{H}}//z$.

  The polarization, calculated for given values of the magnetic field, is substantially underestimated in
comparison with the experimental data. However, this is mainly due to the overestimation of the exchange interactions
in the low-energy model, which makes the magnetic structure harder than in the experiment. If one considers the
slope $P/M$, which is less sensitive to the hardness of the magnetic structure, we will find a
much better agreement with the experiment: for instance, for $\emph{\textbf{H}} // x$ or $y$, the theoretical $P/M$ is about
$50$ $\mu C/(\mu_{\rm B}m^2)$ both for Co$_4$Nb$_2$O$_9$ and Co$_4$Ta$_2$O$_9$, which is comparable with the
experimental value of about $100$ $\mu C/(\mu_{\rm B}m^2)$.\cite{Khanh,Fang2}

\begin{figure}[tbp]
\begin{center}
\includegraphics[width=10cm]{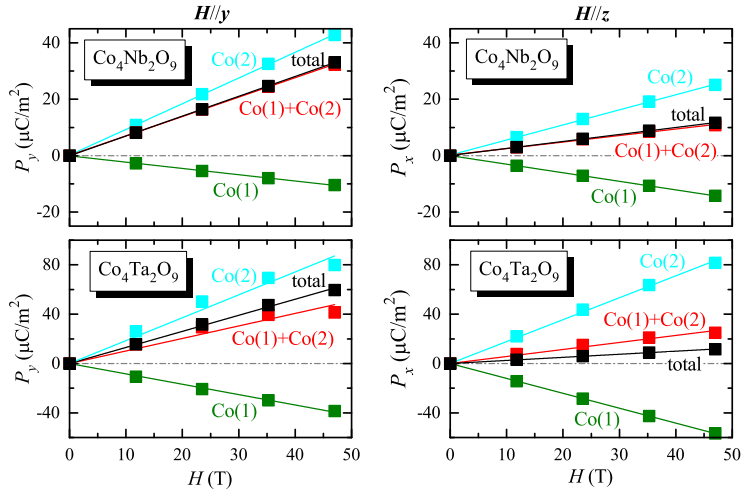}
\end{center}
\caption{(Color online)
Electric polarization in Co$_4$Nb$_2$O$_9$ and Co$_4$Ta$_2$O$_9$ as obtained in the model Hartree-Fock calculations
for the $\boldsymbol{G}_1$ state
when the magnetic field was applied to either Co(1) or Co(2) sublattice. The sum of these two
contributions is denoted as `Co(1)$+$Co(2)', and `total' is the polarization
in the uniform magnetic field applied simultaneously to both magnetic
sublattices.}
\label{fig.PHp}
\end{figure}

  In the $P\overline{3}c1$ structure of Co$_4$Nb$_2$O$_9$ and Co$_4$Ta$_2$O$_9$, the are two inequivalent sublattices
of the Co sites, each of which can exhibit the ME effect: namely, the external magnetic field breaks
the inversion symmetry in each of the two sublattices and, therefore, the total polarization is the
superposition
of such effects in the two sublattices (as well as the interaction between the sublattices). In order to evaluate the
contribution of the magnetic inversion symmetry breaking in each of the sublattices, we apply the nonuniform
magnetic field, acting on either Co(1) or Co(2) sublattices and evaluate the electric polarization.
The results of these calculations for the state $\boldsymbol{G}_1$ are presented in Fig.~\ref{fig.PHp}
(the results for the state $\boldsymbol{G}_2$ are very similar and not shown here).
The most interesting aspect of these calculations is that the electric polarization induced by magnetic field
in the Co(1) and Co(2) sublattices is of the opposite sign.
For Co$_4$Nb$_2$O$_9$, the sum of these two contributions is very close to the
total polarization, calculated in the uniform magnetic field. The small deviation in the case of Co$_4$Ta$_2$O$_9$
is caused by the additional deformation of the magnetic structure in the nonuniform field as well as
stronger inter-sublattice interaction due to the smaller band gap (Fig.~\ref{fig.DOSHF}).
Thus, we find a strong cancelation of contributions of the magnetic sublattices Co(1) or Co(2)
to the electric polarization. In the $\boldsymbol{G}_1$ state, this cancelation is especially strong
for $\emph{\textbf{H}}//z$, which explains smaller value of the induced polarization than for $\emph{\textbf{H}}//y$.
In principle, such effect offers a possibility to control and reverse the electric polarization.

\section{\label{sec:conc}Discussions and Conclusions}
We have clarified the origin of ME effect in the centrosymmetric trigonal systems Co$_4$Nb$_2$O$_9$ and Co$_4$Ta$_2$O$_9$.
Both compounds form the AFM structure, in which the FM chains of Co atoms are antiferromagnetically coupled in the
hexagonal plane. The magnetocrystalline anisotropy tends to align the magnetic moments in the hexagonal plane, thus, lowering
the original $P\overline{3}c1$ space group symmetry. Nevertheless, the magnetic alignment obeys the $\hat{I}\hat{T}$ symmetry,
meaning that, in the ground state, these compounds exhibit neither the net magnetization nor the spontaneous polarization,
but both of them can be induced by either electric or magnetic field, which breaks $\hat{I}\hat{T}$.
In this sense, the situation is similar to the canonical ME compound Cr$_2$O$_3$.\cite{Dzyaloshinskii}
The new aspect of Co$_4$Nb$_2$O$_9$ and Co$_4$Ta$_2$O$_9$ is the existence of two inequivalent Co sublattices,
which contribute to the ME effect. We have found that these contributions are of the opposite signs and, therefore,
partly compensate each other.
Nevertheless, under certain conditions, this balance can be shifted in either way,
thus, giving a possibility to control the direction and magnitude of the ME effect.

  Summarizing results of our joint experimental and theoretical studies, we first note that, as far as
magnetic properties are concerned, Co$_4$Ta$_2$O$_9$ seems to be softer than Co$_4$Nb$_2$O$_9$. Experimentally,
it is clearly observed in the behavior of N\'eel and absolute values of Weiss temperatures, which are systematically
lower in Co$_4$Ta$_2$O$_9$. Moreover, the direct comparison of the behavior of magnetization,
which was reported in Refs.~\onlinecite{Fang1} and \onlinecite{Fang2}, suggests that the AFM structure can be easier deformed
by the magnetic field to induce larger net magnetization in Co$_4$Ta$_2$O$_9$. These experimental data were qualitatively explained by
theoretical calculations of interatomic exchange interactions, which are generally weaker in Co$_4$Ta$_2$O$_9$.
This behavior in turn nicely correlates with the details of the electronic structure calculations of Co$_4$Nb$_2$O$_9$ and Co$_4$Ta$_2$O$_9$.
The quantitative differences
between the theory and experiment
are related to the fact that the theoretical calculations were performed using
minimal effective Hubbard-type model, constructed only for the Co$3d$ states, which overestimates the tendencies
towards the antiferromagnetism.\cite{Ba2CoGe2O7,CrO2}

  From the viewpoint of the minimal electron model considered in the present work the main factor controlling the
behavior of the polarization in Co$_4$Nb$_2$O$_9$ and Co$_4$Ta$_2$O$_9$ should be the softness of the magnetic structure and
its ability to be deformed by the external magnetic filed. Then, we would expect that the application of the magnetic filed
should induce larger polarization in magnetically softer Co$_4$Ta$_2$O$_9$ than in Co$_4$Nb$_2$O$_9$,
as it was indeed obtained in our theoretical calculations. However, there is
number of experimental data, which suggest the opposite tendency. Particularly, the dielectric response to the magnetic
field near $T_{\rm N}$, studied in the present work, is weaker in Co$_4$Ta$_2$O$_9$. Moreover, the experimental polarization,
induced for a given magnetic field is systematically smaller in Co$_4$Ta$_2$O$_9$ than in Co$_4$Nb$_2$O$_9$.\cite{Fang1,Fang2}
Yet, the experimental situation is somewhat controversial because the first direct measurements of the ME susceptibility
suggested the opposite tendency:\cite{Fischer} the susceptibility was systematically larger in Co$_4$Ta$_2$O$_9$,
but exhibited some nonmonotonous behavior as the function of temperature. Thus, we believe that this issue requires a more systematic study
and it would be important,
for instance, to measure directly the ME susceptibility for the single crystalline sample.

  Below, we discuss some factors, which have not been taken into account by our theoretical model and which can alter some
of our conclusions and also affect the comparison with the experimental data.

  (i) The magnetostriction effect in the ordered magnetic phase can play an important role. In our theoretical calculations
we used the experimental structure parameters, measured around the room temperature: $T$ = 297 K for Co$_4$Nb$_2$O$_9$ and
$T$ = 298 K for Co$_4$Ta$_2$O$_9$.\cite{Co4Nb2O9str,Co4Ta2O9str} These parameters do not take into account some possible change
of the crystal structure, which may occur below $T_{\rm N}$. Indeed, since the magnetic alignment in the hexagonal plane lowers
the original $P\overline{3}c1$ symmetry, it is reasonable to expect also some changes in the crystal structure, which adjust
to the change of the magnetic structure. In this sense, it is somewhat surprising that no structural phase transition
has been observed in the experiment.\cite{Khanh} Nevertheless, some structural change below $T_{\rm N}$ cannot be completely
ruled out because of the following observations. First, the dielectric constant in Co$_4$Nb$_2$O$_9$ exhibits a clear
upturn below $T_{\rm N}$, even without magnetic field, as it was observed in Ref.~\onlinecite{Xie} and also confirmed by our measurements.
As it was argued in Ref.~\onlinecite{Xie}, this change can be
of magnetostrictive origin. Second, the electric polarization induced by the magnetic field in Co$_4$Nb$_2$O$_9$
has a pronounced off-diagonal component.\cite{Khanh} This finding is inconsistent with the $P\overline{3}c1$ symmetry, according to
which the polarization should be parallel to the $y$ axis for $\emph{\textbf{H}} // x$ or $y$ (depending on the magnetic state), or
parallel to either $y$ or $x$ axis when $\emph{\textbf{H}} // z$, but there should be no off-diagonal components of the polarization
in the $xy$ plane.

\begin{figure}[tbp]
\begin{center}
\includegraphics[width=10cm]{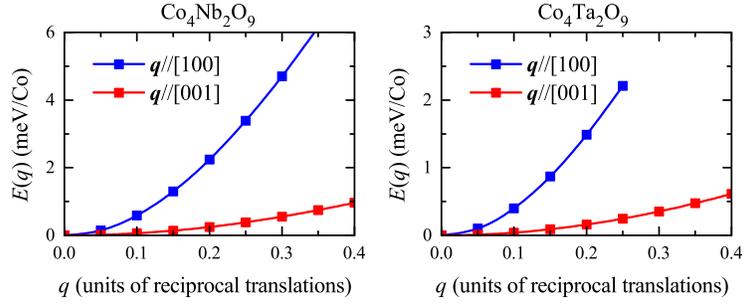}
\end{center}
\caption{(Color online)
Dependence of the total energy on the spin-spiral vector $q$ as obtained in the mean-field Hartree-Fock calculations
for Co$_4$Nb$_2$O$_9$ and Co$_4$Ta$_2$O$_9$.}
\label{fig.Eq}
\end{figure}

  (ii) Another important issue is the possible change of the magnetic structure, which can be induced, for instance, by
pooling electric field used in some of the experiments (e.g., in Ref.~\onlinecite{Khanh}) or some other factors.
How robust is the obtained AFM ground state and whether Co$_4$Nb$_2$O$_9$ and Co$_4$Ta$_2$O$_9$
are the conventional ME systems? Or, can these compounds under certain conditions become type-II multiferroics,
where the onset of electric polarization is triggered by some massive changes in the magnetic structure, which breaks spontaneously the inversion symmetry?\cite{MF_review} The above scenario looks quite feasible taking into account the complexity of magnetic interactions (Fig.~\ref{fig.J}),
many of which are antiferromagnetic, not necessarily restricted by the nearest neighbors, and competing with each other.
On many occasions such behavior is responsible for the type-II multiferroism.\cite{MF_review,typeII}
In order to explore this possibility in Co$_4$Nb$_2$O$_9$ and Co$_4$Ta$_2$O$_9$ we have performed self-consistent HF spin-spiral calculations without the SO coupling, based on the generalized Bloch theorem.\cite{Sandratskii} The results are presented in Fig.~\ref{fig.Eq}.
In this case, $\boldsymbol{q}=0$ correspond to the ground-state AFM alignment,
which is deformed for finite spin-spiral propagation vectors $\boldsymbol{q}$.
For each value of $\boldsymbol{q}$, the magnetic moments in the $xy$ plane were allowed to freely relax in order to minimize
the total energy of the system. For $\boldsymbol{q} // z$ the dependence $E(q)$ is very
flat (contrary to $\boldsymbol{q} // x$) when the energy change remains less than 1 meV/Co
even for relatively large $q=|\boldsymbol{q}|$. In such situation the ground state is still $\boldsymbol{q}=0$.
However, any perturbation of the magnetic system, linear in $\boldsymbol{q}$, can induce the transition to a
noncollinear state with the broken inversion symmetry, which will further affect the behavior of electric polarization.
For instance, such transition can be caused by the electric field, leading to the off-centrosymmetric atomic displacements
and appearance of Dzyaloshinskii-Moriya interactions,\cite{DM} connecting different unit cells.

\textit{Acknowledgements}.
The work of IVS is partly supported by the grant of
Russian Science Foundation (project No. 14-12-00306). TVK was supported by Grant-in-Aid for Scientific Research C 26400323 from JSPS.


\begin{thebibliography}{99}
\bibitem{request}
All parameetrs of the model Hamiltonian are available upon request.
\end{thebibliography}
\end{document}